\documentclass[aps,prd,preprint,floatfix]{revtex4}
\usepackage{epsfig}
\usepackage{latexsym}

\usepackage{epsfig}%
\usepackage{longtable}
\usepackage{eucal}
\begin{document}
\setlength{\topmargin}{-0.25in}
\title{Large Field Cutoffs in Lattice Gauge Theory}


\author{L. Li and Y. Meurice}
\thanks{Talk presented by Yannick Meurice (yannick-meurice@uiowa.edu) at the Workshop on QCD in Extreme
Environments, Argonne National Laboratory, 29th June to 3rd July, 2004.}
\affiliation{Department of Physics and Astronomy\\ The University of Iowa\\
Iowa City, Iowa 52242 \\ USA
}


\date{\today}

\begin{abstract}

In pure gauge $SU(3)$ 
near $\beta \simeq 6$, weak and strong coupling expansions 
break down and the MC method
seems to be the only practical alternative. We discuss the possibility 
of using a modified version of perturbation theory which relies on a large field cutoff 
and has been successfully applied to the double-well potential (Y. M., PRL 88 141601).
Generically, 
in the case of scalar field theory, the weak coupling expansion is unable to reproduce the exponential suppression of the large field configurations. This problem can be solved by introducing a large field cutoff $\phi _{max}$. The value of $\phi_{max}$ can be chosen 
to reduce 
the discrepancy with the original problem. This optimization can be approximately 
performed using the strong coupling expansion and bridges the gap between 
the two expansions.
We report recent attempts to extend this procedure 
for $SU(3)$ gauge theory on the lattice. 
We compare gauge invariant and gauge dependent (in the Landau gauge) criteria to sort the configurations into ``large-field'' and ``small-field'' 
configurations. 
We discuss the convergence of lattice perturbation theory and the 
way it can be modified in order to obtain results similar to the scalar case.

\end{abstract}


\maketitle

\section{Introduction}

A common challenge for quantum field theorists consists in 
finding accurate methods in regimes where existing expansions break down. 
In the RG language, this amounts to find acceptable interpolations for the RG flows in intermediate regions between fixed points.
In the case of scalar field theory, the weak coupling expansion is unable to reproduce the exponential suppression of the large field configurations operating 
at strong coupling. This problem can be cured by 
introducing a large field cutoff $\phi _{max}$ which eliminates Dyson's instability.
One is then considering a slightly different problem, however a judicious choice of $\phi _{max}$ can be used to reduce or eliminate the discrepancy with the original problem (i. e., the problem with no field cutoff). 
This optimization procedure can be approximately 
performed using the strong coupling expansion and naturally bridges the gap between 
the weak and strong coupling expansions. 

The Workshop on QCD in Extreme
Environments was held right after the Lattice 2004 conference. 
The talk of K. Wilson about the early days of lattice gauge theory 
was a very inspirational moment of Lattice 2004. He 
stressed the importance of, in his own words, ``butchering field theory'' in the development of the 
RG ideas and recommended that we keep doing it. 
In the following, we will be butchering field theory in the space of field configurations.
We are interested in the effects of a large field cutoff on observables (we expect the effect to be small 
if the field cutoff is large enough and the observable is not a product of too many fields) and on the 
coeffients of the perturbative series for the observables (we expect the effect to be drastic for the 
large order behavior).
For scalar fields, there are many ways to accomplish this task.
The configurations can be ranked according 
to the largest 
absolute value of the field  or according to the average over the sites of even powers of the field. The larger the  power is, the more emphasis is put on the 
configurations with the largest field values. As one may suspect, there exists 
correlations among the results obtained with different cutoff procedures. 
For gauge theory, we can define the concept 
of small or large field configurations in the Landau gauge and in a gauge invariant way. 
This was one of the questions that we discussed  at Lattice 2004 and an account can be found in 
Ref. \cite{lat2004}. Instead of duplicating, we will rather give an elementary discussion of our motivations 
and existing results 
in the scalar case and explain how we expect to extend them in the gauge case. Recent progress are briefly 
discussed at the end.


\eject

\section{Basic ideas in the scalar case}

The best way to understand why the perturbative series of $\lambda \phi^4$ problems in various dimensions generically have a zero radius radius of convergence is to consider the integral
\begin{equation}
\int_{-\infty}^{+\infty}d\phi e^{-\frac{1}{2}\phi^2-\lambda \phi^4}\neq \sum_0^{\infty}
\frac{(-\lambda)^l}{l!} \int_{-\infty}^{+\infty}d\phi e^{-\frac{1}{2}\phi^2}\phi^{4l}
\end{equation}
The peak of the integrand of the r.h.s. moves too fast when the order increases.
More precisely,
${\rm e}^{-(1/2)\phi^2}\phi^{4k}$ is maximum at $\phi_k$ such that 
$\phi_k^2=4k$. 
The truncation of 
${\rm e}^{-\lambda \phi^4}$ at
order $k$ is accurate provided that $\lambda \phi^4 <<k$.
A good accuracy in the region where the integrand is 
maximum requires 
$\lambda \phi_k^4 <<k$,
which implies
$\lambda<<{\frac{1}{16 k}}$. Note also that 
the exponential function converges uniformly over a finite interval but {\it not} over $(-\infty,+\infty)$ and consequently one cannot interchange the sum and the integral.

On the other hand, if we introduce a field cutoff, the peak moves outside of the integration range and we get a converging expansion
\begin{equation}
\int_{-\phi_{max}}^{+\phi_{max}}d\phi e^{-\frac{1}{2}\phi^2-\lambda \phi^4}= \sum_0^{\infty}
\frac{(-\lambda)^l}{l!} \int _{-\phi_{max}}^{+\phi_{max}}d\phi e^{-\frac{1}{2}\phi^2}\phi^{4l}
\end{equation}
In general we expect that for a finite lattice, the partition function $Z$ calculated with a field cutoff is convergent and $\ln(Z)$ has a finite radius of convergence. 
The problem with the field cut differs from the original problem but the difference
can be made exponentially small.
The method works well in nontrivial examples.
This has been checked \cite{convpert} for the hierarchical model 
and in the continuum for quantum 
mechanical problems (the anharmonic oscillator and the double-well potential).

\section{Significant Digits versus Coupling}

\begin{figure}[h]
\includegraphics[width=2.7in,angle=0]{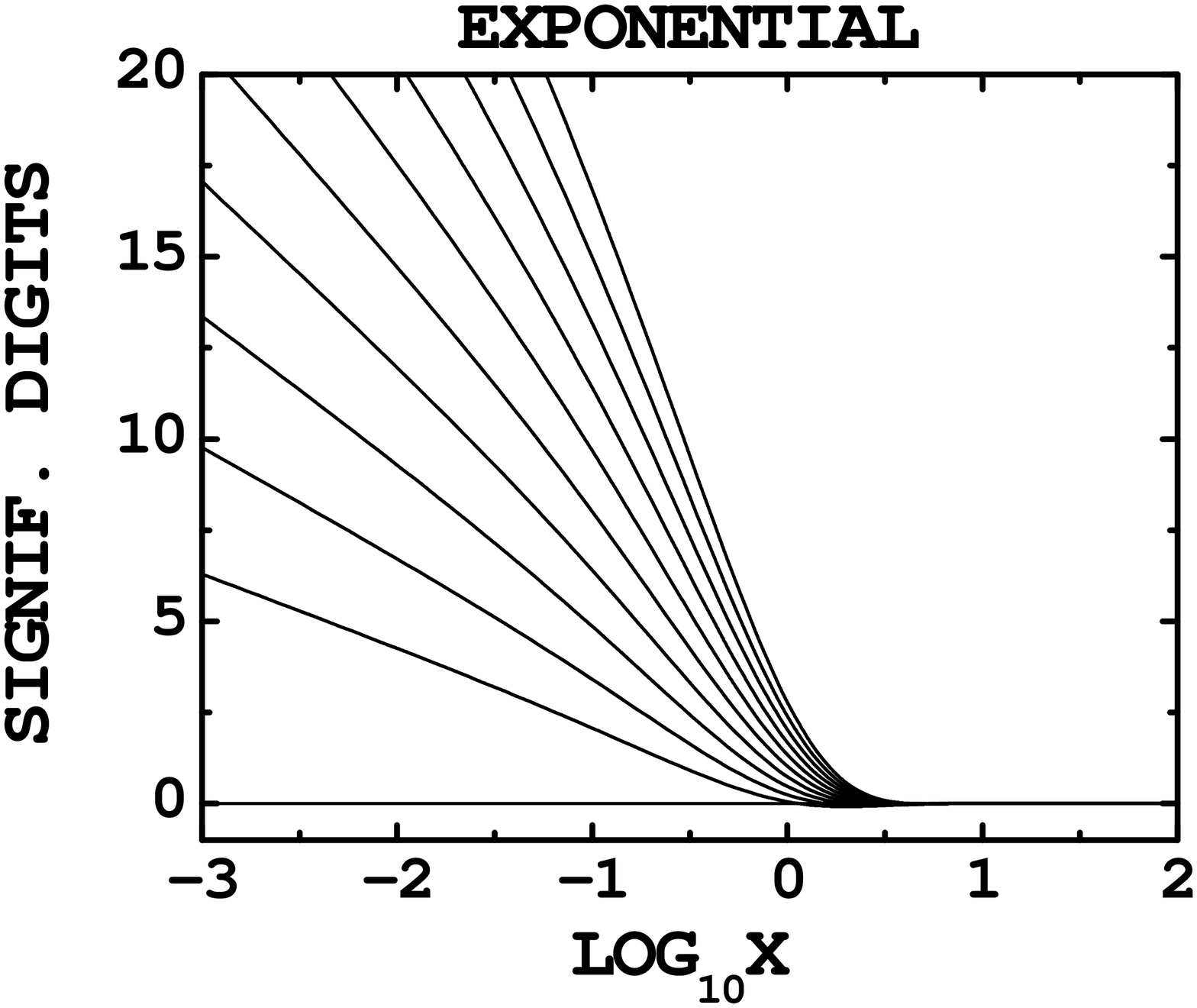}
\includegraphics[width=2.7in,angle=0]{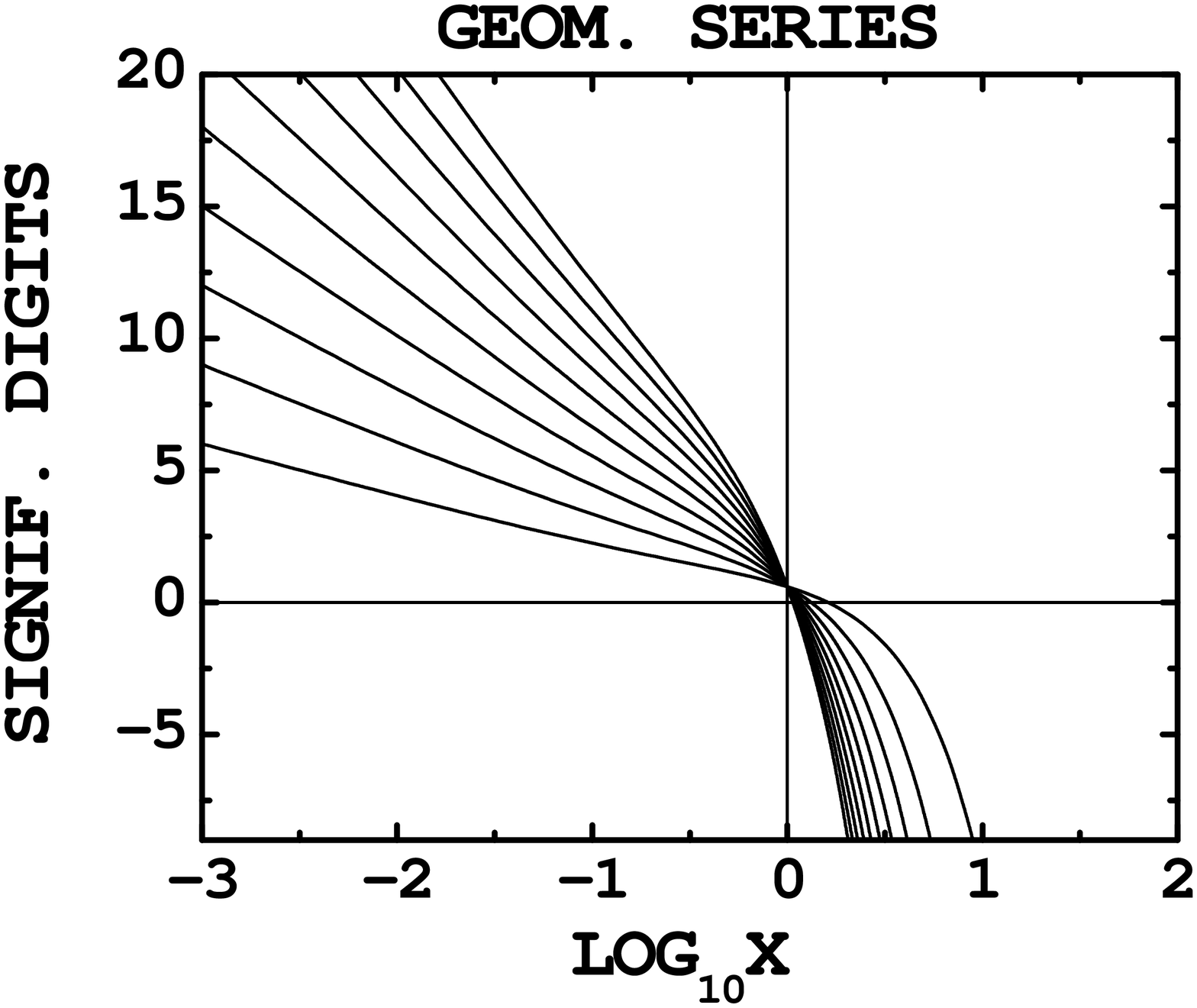}
\includegraphics[width=3.2in,angle=0]{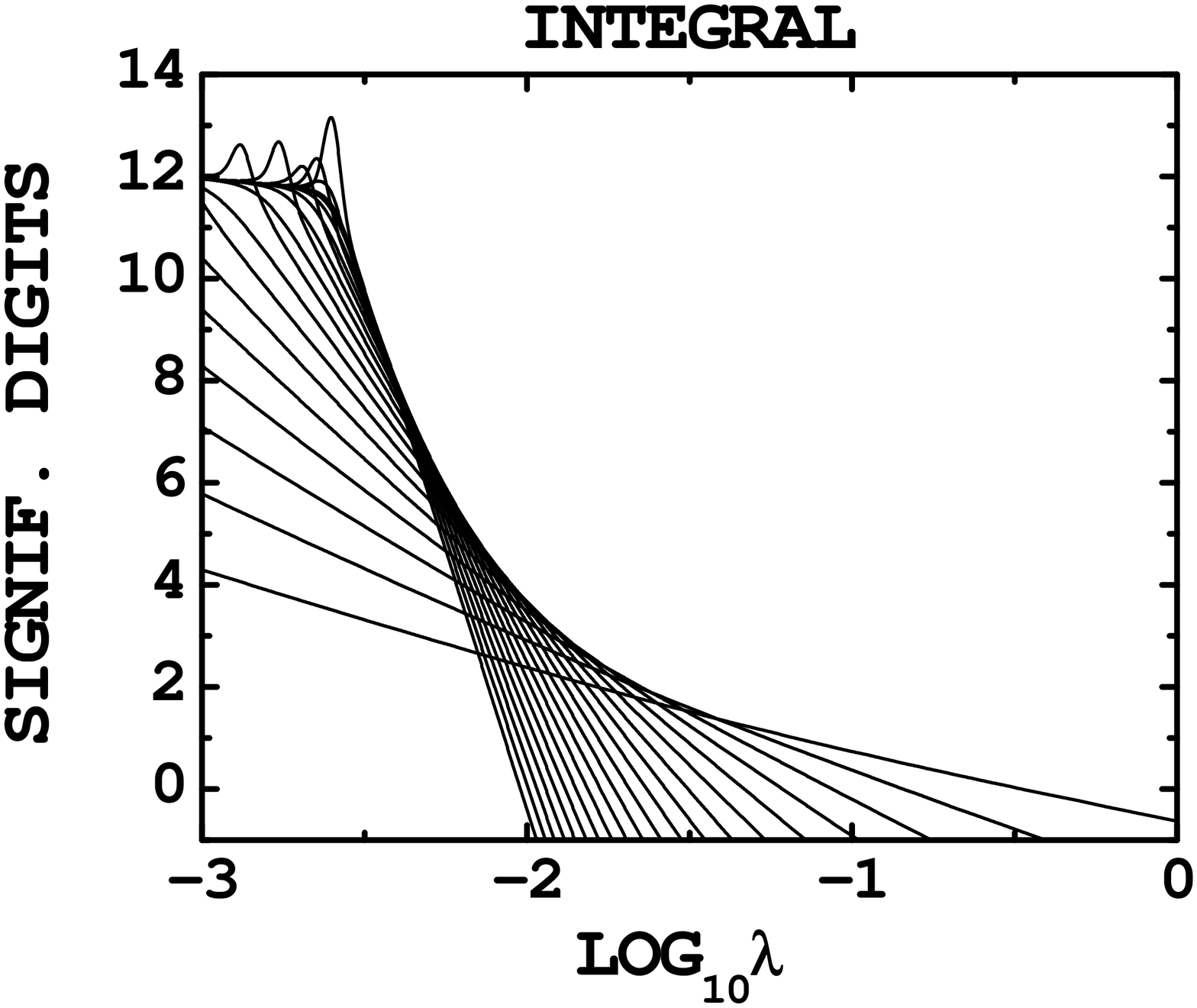}
\caption{Number of (correct) significant digits 
obtained by perturbation theory at order 1, 2, ..., 10 
for ${\rm e}^{x}$ (upper left) and $\frac{1}{1+x}$ (upper right), and 
orders 1, 2, ..., 20 for the integral  discussed in section 2 (lower). As the order increases, the lines ``rotate'' clockwise.}\label{fig:sd}
\end{figure}

In this section, we describe graphical representations of the accuracy reached at different orders in perturbation theory. We typically want to know the number of 
(correct) significant digits that can be obtained at a given order for a given 
coupling. 
\eject
Three situations are represented in Fig. \ref{fig:sd}. For ${\rm e}^x$, the accuracy always increases when we increase the order. In the 
case of $1/(1+x)$, this is the case only if $|x|<1$ and the lines 
have a focus point at $x=1$. On the other hand, for the integral discussed in
the previous section, the lines 
move left as they rotate and one sees an envolope that delimitates a 
`` forbidden region'' for the accuracy of perturbation theory. At fixed and not too large coupling, the accuracy first increases and then decreases with the order. The ``rule of thumb'' consists in stopping when the accuracy is optimal,  
in other words, when we reached the envelope discussed above. Note also that the 
lines flatten near 14 on the left of the graph. This simply reflects that we have 
only 14 digits of accuracy in our numerical calculation of the integral.
\eject

\begin{figure}[h]
\centerline{\psfig{figure=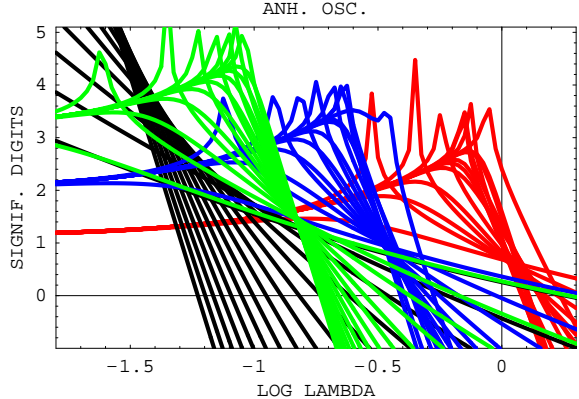,width=3in}}
\vskip10pt
\caption{Number of significant digits at order 1, 2, 3...., 15 ,
obtained with regular perturbation 
theory (black) and
with $\phi_{max}$ = 3 (green), 2.5 (blue) and 2 (red),
as a function of $\lambda$, 
for the ground state of the anharmonic oscillator.}
\label{fig:cutanh}
\end{figure}

When a field cut is introduced, the series apparently become convergent and we 
can make a significant, but localized in coupling, incursion in the forbidden 
region of perturbation theory. This is illustrated 
for three different field cuts in Fig. \ref{fig:cutanh} 
in the case of the ground state of the anharmonic oscillator.

In an ideal world, we would pick a field cut large enough to reduce the errors 
(due to the field cut) to an acceptable level. We would then calculate enough terms 
in order to reach an accuracy consistent with this level.
In practice, we are usually limited to calculations up to a certain order.
The field cut can then be chosen in order to minimize the 
error at that order. From Fig. \ref{fig:cutanh}, one can see that it is possible to pick a 
cut that makes the accuracy optimal in the neighborhood of some given value of the 
coupling $\lambda$. When the numerical answer is known, it is easy to adjust the 
field cut in order to optimize the accuracy. When the answer is not known, one can use an approximation. In Ref. \cite{optim}, we showed that for the integral Eq. (1), 
the strong coupling expansion can be used to determine approximately the optimal field cut. This way 
strong and weak coupling expansions can be combined coherently.

Up to now, we have defined the field cut locally in configuration space (at each 
lattice site). It is however possible to proceed differently and to use the average of even powers of the fields to sort the configurations.
There exist correlations among these indicators \cite{lat2004}. 

\section{Lattice Perturbation Theory}
\begin{figure}
\centerline{\psfig{figure=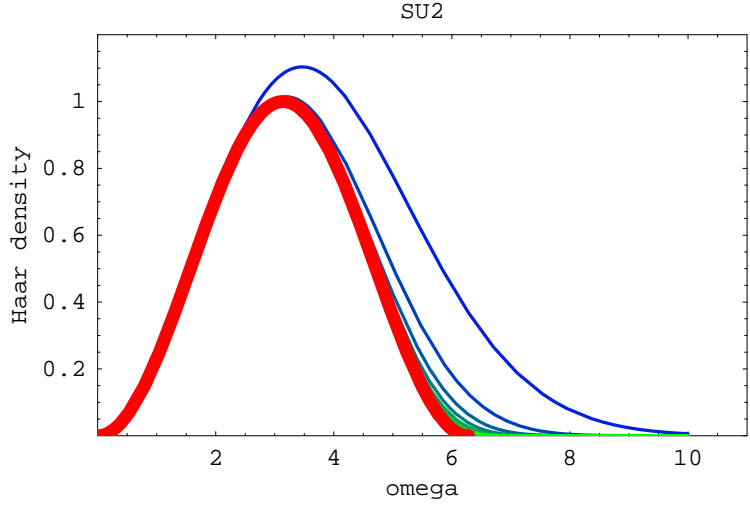,width=3.9in}}
\caption{The red curve is the Haar density $\sin^2(\omega/2)$ with $\omega\leq 2\pi$.
The blue/green curves represent the approximations $(\omega/2)^2e^{-\frac{\omega^2}{12}-\frac{\omega^4}{1440}-\dots}$. As the order increases, the curve gets more green.}
\label{fig:haar1}
\end{figure}

\begin{figure}
\vskip30pt
\centerline{\psfig{figure=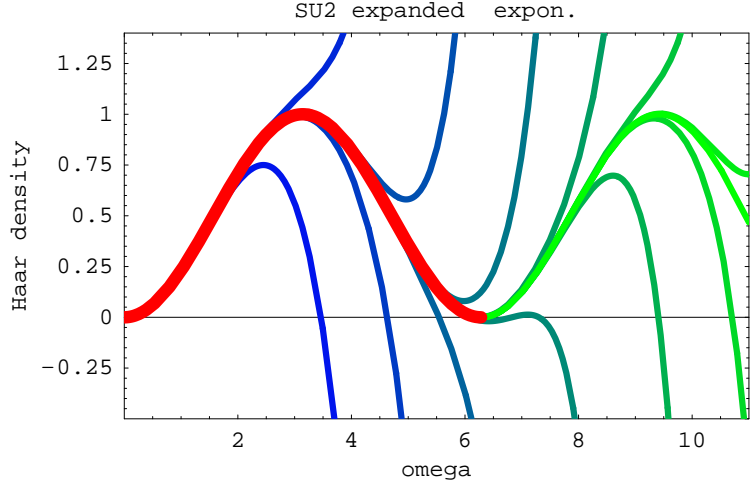,width=3.9in}}
\caption{Same as Fig. \ref{fig:haar1} but with 
$(\omega/2)^2e^{-\frac{\omega^2}{12}-\frac{\omega^4}{1440}-\dots}$ replaced by its expansion 
$(\omega/2)^2(1-\frac{\omega^2}{12}-\dots)$.
The radius of convergence is infinite and the tails are restored.}
\label{fig:haar2}.
\end{figure}

We now report our attempts to extend the modification of perturbation theory discussed above in 
the scalar case to LGT. Perturbation theory for LGT has been developed almost 20 years ago \cite{heller}.
Exact calculations up to 3 loops \cite{alles93} and numerical calculations for 8 \cite{direnzo95} and 10 loops \cite{direnzo2000} 
are available. It proceeds in 3 steps. 
\begin{enumerate}
\item
With the convention $\beta=2N/g^2$, we set $U=e^{igA}$ at every link.
\item
We extend the range of integration OF $A$ to $R^{N^2-1}$ (anything else would be unpratical!)
\item
We then expand in powers of $g$
\end{enumerate}
In step 2, we added the integration ``tails''. This presumably makes the series nonconvergent (asymptotic) as 
one can observe in the case of large argument expansions of Bessel functions. In step 3, one needs to expand the 
Haar measure in power of $g$. As the original Haar measure is compact and provides a natural field cut, we would 
like to see what happens when the integral get decompactified in step 2. For this purpose, we consider the simple 
example of a one link $SU(2)$ integral.
We use the parametrization of $SU(2)$
\begin{equation}
U=\exp(\frac{i}{2}\vec{\tau}.\vec{\omega})=\cos(\omega/2)+i\vec{\tau}.\hat{\omega}\sin(\omega/2)
\end{equation}
with $\hat{\omega}$ covering the 2-sphere and $0\leq\omega\leq 2\pi$. With this parametrization we cover the 
$SU(2)$ manifold exactly once. Alternatively, we could extend the range to $4\pi$, but then we need to identify 
opposite points on the sphere if we want to avoid a double coverage.  
In these coordinates, the invariant Haar measure reads
\begin{equation}
dU=\frac{d\omega}{2\pi}\sin^2(\omega/2)\frac{d^2\hat{\omega}}{4\pi}=
\frac{d^3\omega}{32\pi^2}\left(\frac{\sin(\omega/2)}{\omega/2}\right)^2=
\frac{d^3\omega}{32\pi^2}e^{-\frac{\omega^2}{12}-\frac{\omega^4}{1440}\dots}
\end{equation}
After we switched to the $R^3$ measure $d^3\omega$, we obtain the Haar correction 
${\rm e}^{-Log[(\frac{\sin(\omega/2)}{\omega/2})^2]}$. Note that $Log[(\frac{\sin(\omega/2)}{\omega/2})^2]$ has a radius of convergence $2\pi$, but all the coefficients are negative. So when we expand
{\it in the exponential}, the large negative contributions in the region $\omega>2\pi$ effectively cutoff 
these contributions. This is illustrated in Fig. \ref{fig:haar1}.
However, in perturbation theory we have $\vec{\omega}\equiv g\vec{A}$ and we need to expand the exponential
in powers of $g$. The measure then ``periodicizes'' and we obtain a multiple coverage of the manifold as
shown in Fig. \ref{fig:haar2}. The logarithm of the Haar measure is also used in the context of gluon 
equations of state \cite{mo}.
\eject
\vskip10pt
\vskip30pt
\eject
\section{Comparison with the double-well potential}
\begin{figure}[h]
\vskip30pt
\includegraphics[width=2.9in,angle=0]{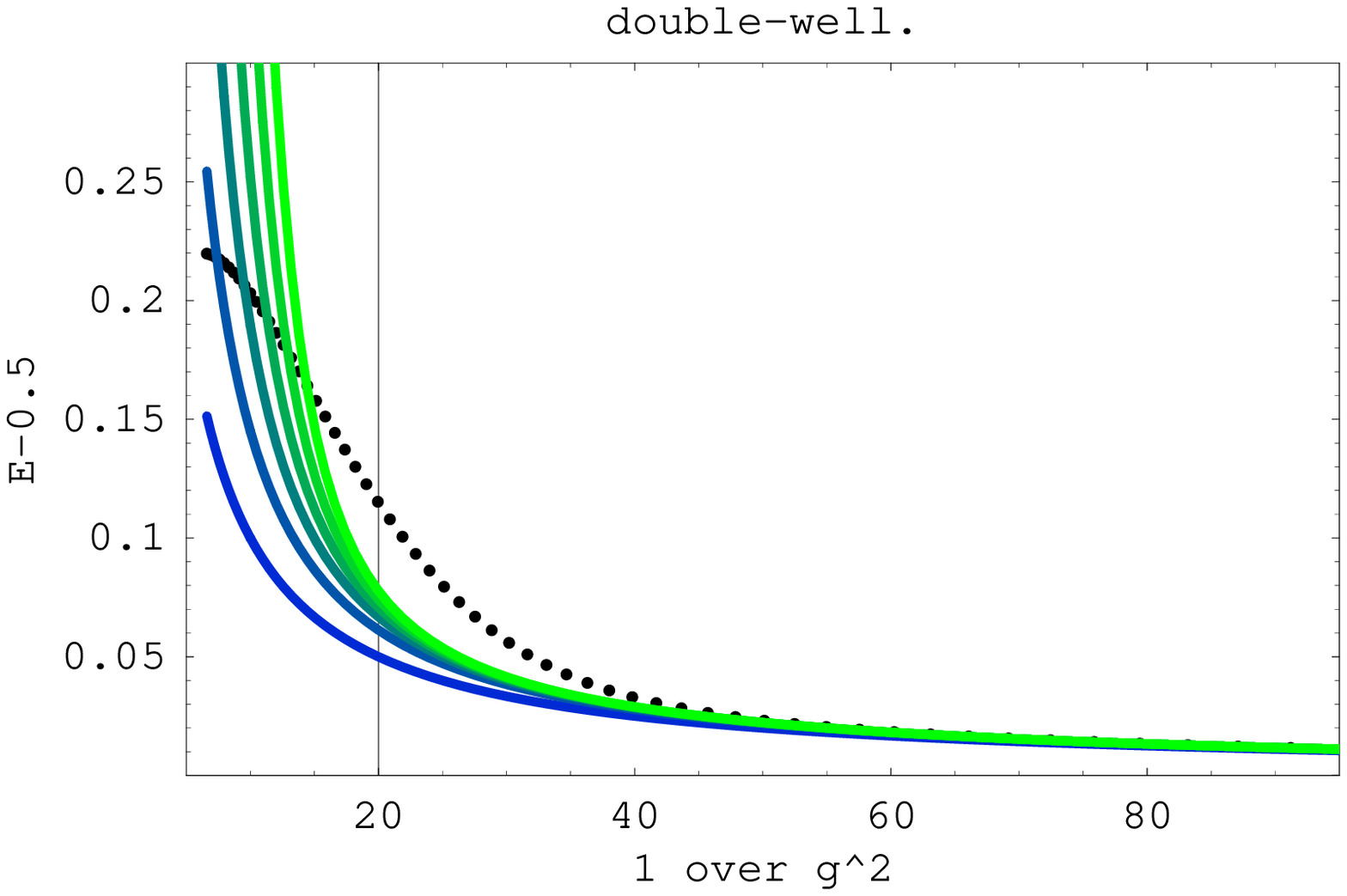}%
\hskip0.5in
\includegraphics[width=2.8in,angle=0]{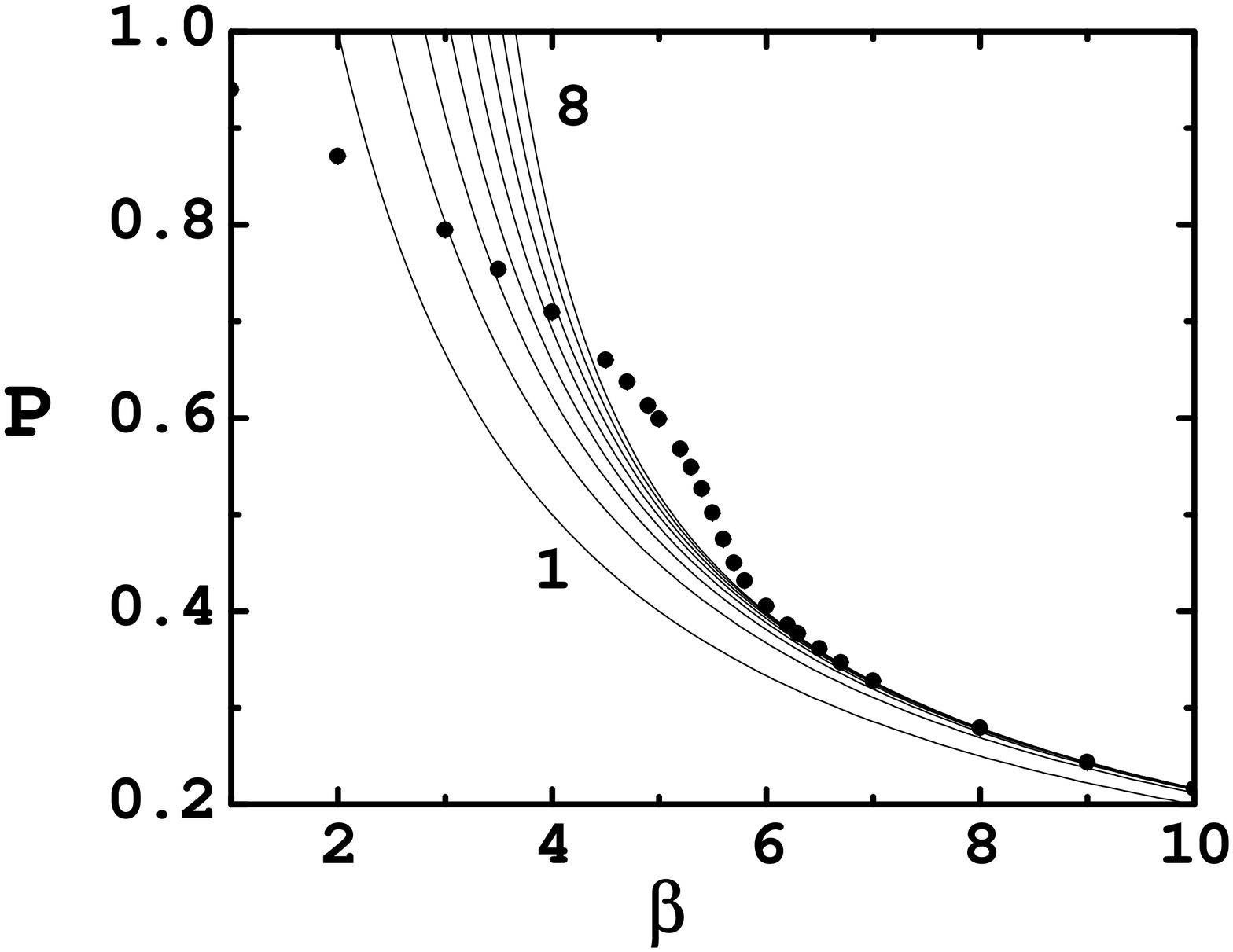}%
\caption{$E_0-(1/2)$ for the double well potential (left). The blue/green lines represent the first orders of perturbation theory. As the order increases, the curve gets more green. 
The dots represent the numerical values.
A similar graph (except for the colors) for $P$ in $SU(3)$ is shown on the right.}
\label{fig:num}
\end{figure}

\begin{figure}[ht]
\includegraphics[width=3in,angle=0]{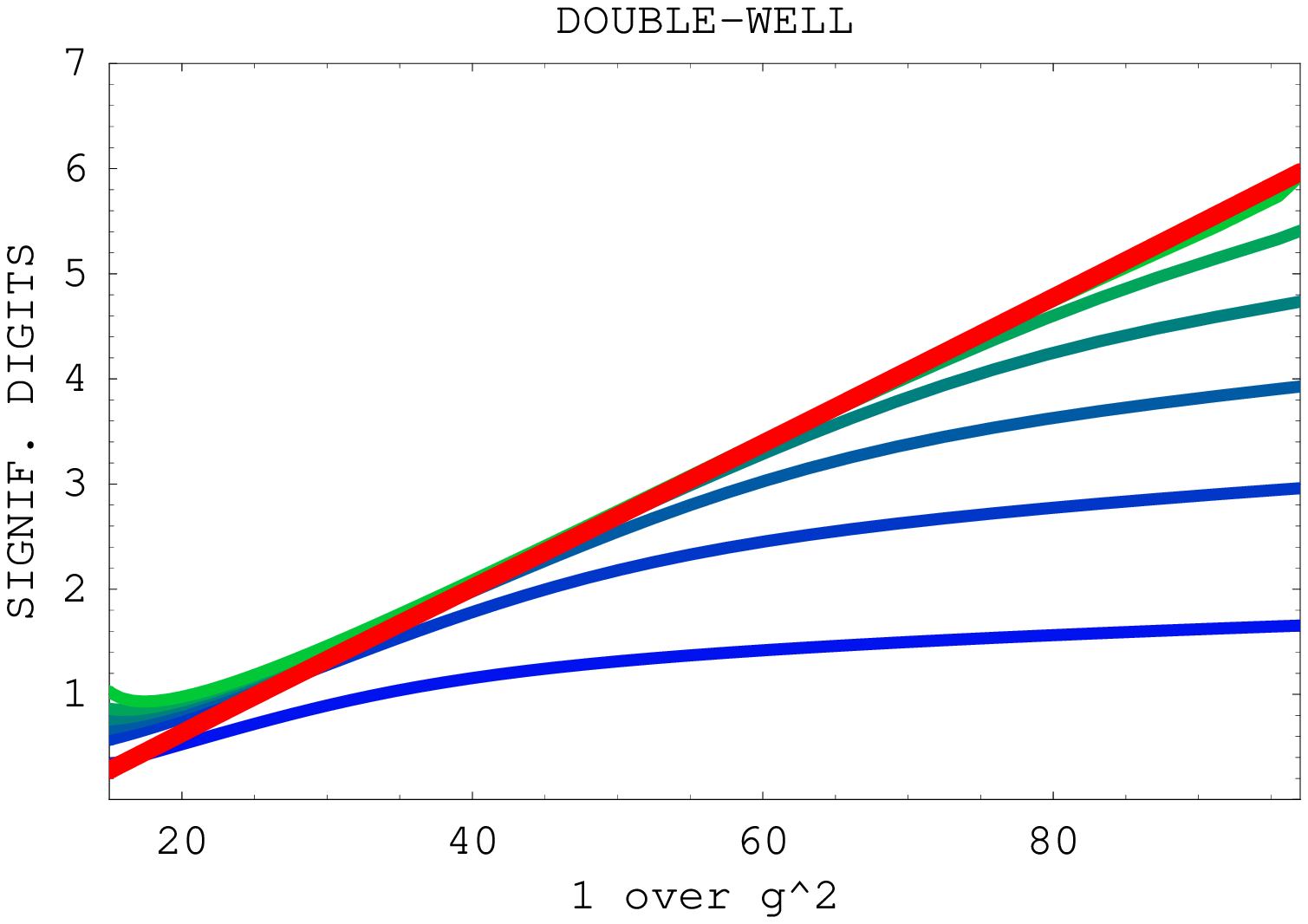}%
\hskip0.4in
\includegraphics[width=3in,angle=0]{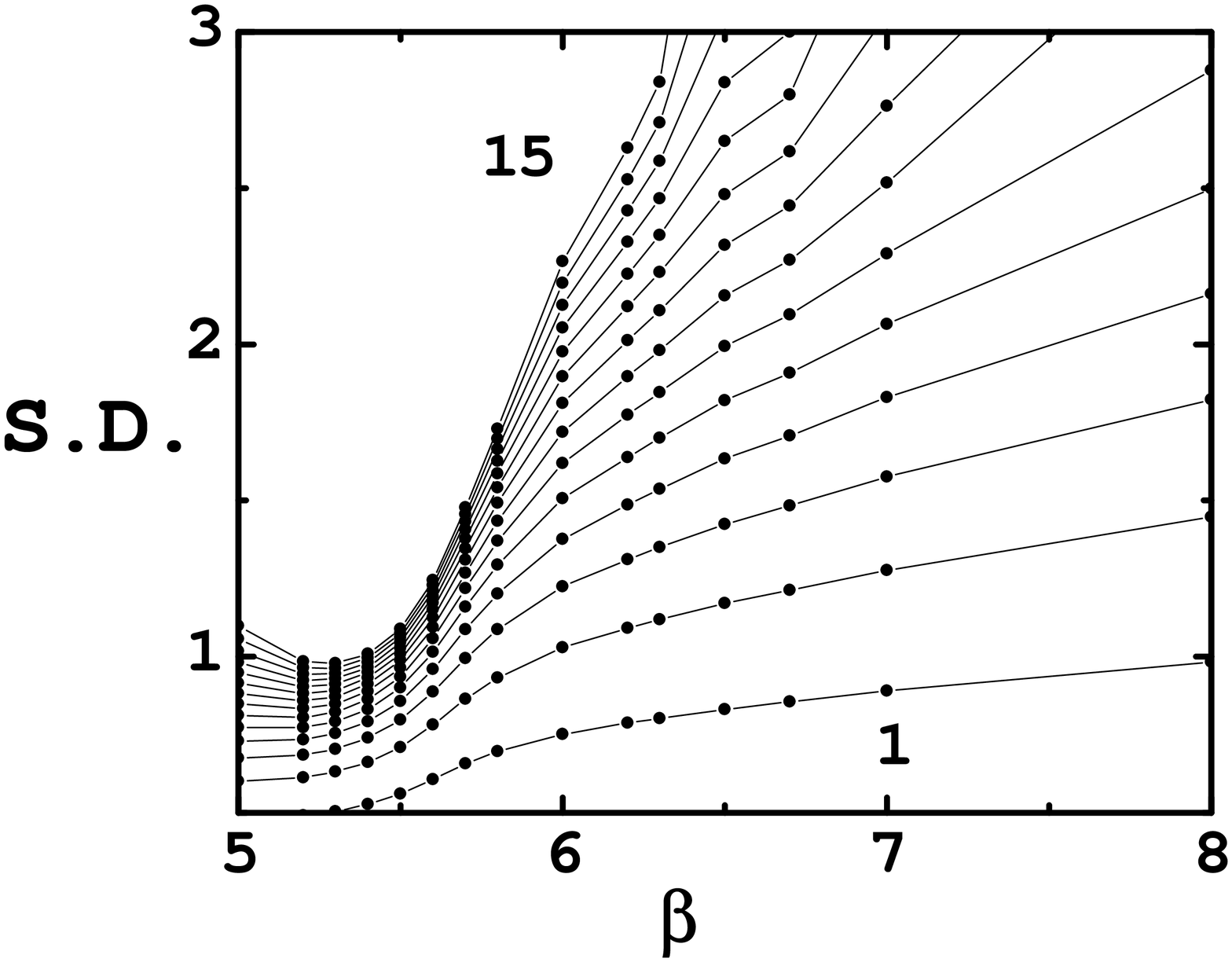}%
\caption{Significant digits for succesive orders in perturbation theory corresponding to 
the cases displayed in Fig. 
\ref{fig:num}. For the double-well (left), 
the significant digits are bounded by the instanton effect 
$\delta E_0=\frac{1}{g\sqrt{\pi}}\exp(-\frac{1}{6g^2})$ (red curve).}
\label{fig:acc}
\end{figure}

\begin{figure}[ht]
\vskip20pt
\centerline{\psfig{figure=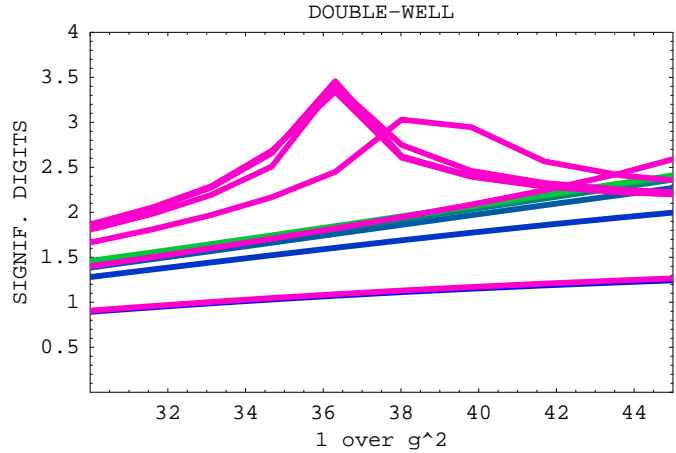,width=3.5in}}
\caption{Significant digits of $E_0-(1/2)$ for the double well potential. The blue/green lines represent the first orders of perturbation theory as in Fig. \ref{fig:acc}. 
The purple lines represent the accuracy for a perturbative series calculated with a cut.}
\label{fig:dwcut}
\end{figure}

\begin{figure}[ht]
\centerline{\psfig{figure=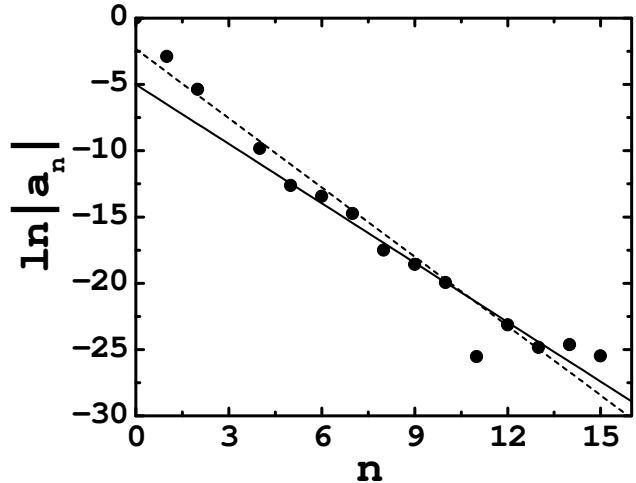,width=3.8in}}
\caption{Logarithm of the absolute value of the coefficients of the strong coupling expansion of $P$. 
The two linear fits done with different subsets of points, correspond to a radius of convergence of 4.45 and 5.71 repectively}
\label{fig:strong}
\end{figure}

The ground state of the double-well potential 
\begin{equation}
V(y)=(1/2)y^2-gy^3+(g^2/2)y^4
\end{equation}
can be expanded in powers of $g^2$. Except for the zeroth order contribution, all the coeffients of the series are negative and their magnitude grow factorially with 
the order. The Borel transform has poles on the positive real axis. The difference between 
the beginning of the perturbative series and the numerical values 
is bounded by the instanton effect 
\begin{equation}
\delta E_0=\frac{1}{g\sqrt{\pi}}\exp(-\frac{1}{6g^2})\ .
\end{equation}
Qualitatively similar features are expected for the perturbative expansion of $P$ in pure gauge $SU(N)$ defined as 
\begin{equation}
\label{eq:pdef}
P(\beta)\equiv (1/\mathcal{N}_p)\left\langle \sum_p
(1-(1/N)Re Tr(U_p))\right\rangle \ ,
\end{equation}
with 
\begin{equation}
\mathcal{N}_p\equiv\ L^D D(D-1)/2\ .
\end{equation}
The comparison between the numerical values and successive orders are shown in 
Fig. \ref{fig:num}. 
The accuracy of successive orders in perturbation theory are shown in Fig. \ref{fig:acc}. Note that 
unlike the scalar case, the weak coupling (large $\beta$) is now displayed on the right of the figure.
Appropriate field cuts can restore the instanton effects in the perturbative series \cite{convpert}. 
This is illustrated in Fig. \ref{fig:dwcut} where the modified series allows us to go above the instanton 
envelope. 
\eject
We expect to be able to achieve similar results for $SU(3)$. In particular, we expect to be able 
to use the strong coupling expansion to obtain an optimal choice of field cut, since the validity of this expansion seems to extend close to the scaling window. Fig. \ref{fig:strong} indicates that the 
radius of convergence of the strong coupling expansion is between 4 and 6 for $\beta$. This series was calculated using the expansion of the free energy of Ref. \cite{balian74err}.

\eject
\section{Work in progress and recent developments}
At the time of the workshop, we presented results related to the questions discussed below. Since our understanding has evolved in the meantime, we will give a brief summary and refer to recent preprints 
for more details.
\subsection{Field cuts in LGT}

We have attempted to follow the same procedure as for the scalar models, for gauge models using the Landau gauge where $1-(1/N)ReTrU_{link}$ should play a role analogous to $\phi^2$ in
scalar models. We found correlations between the lattice average of this quantity and the average action. However, we found no correlations between the 
average and the maximum value. 
These results are explained in more detail in the Proceedings of Lattice 2004 \cite{lat2004}.
The lack of correlation is due to the imperfect way the 
Landau gauge condidtion is implemented numerically. This is being remedied \cite{lilipro}.
\subsection{Gluodynamics at negative $g^2$} 
We considered Wilson's $SU(N)$ lattice gauge theory (without fermions) at negative 
values of $\beta= 2N/g^2$ and for $N$=2 or 3. We showed that in the limit $\beta 
\rightarrow -\infty$, the path integral is dominated by configurations where links variables are set to 
a nontrivial element of the center on selected non intersecting lines. For $N=2$, these configurations can be characterized by a unique gauge invariant set of variables, while for $N=3$ a multiplicity growing with 
the volume as the number of configurations of an Ising model is observed. In general, there is a discontinuity in the 
average plaquette when $g^2$ changes its sign which prevents us from having a convergent series in $g^2$ 
for this quantity.
For $N=2$, a change of variables relates the gauge invariant observables at positive and negative values of $\beta$. 
For $N=3$, we derived an identity relating the observables at $\beta$ with those at $\beta$ rotated by $\pm 2\pi/3$ in the complex plane and showed numerical evidence for a Ising like first order phase transition near $\beta=-22$. 
So far we see no obvious connections to the known singularities \cite{kogut80}.
These results are discussed in more detail in a recent preprint \cite{negbet}. For another approach of problems at negative coupling see Ref. \cite{bender}.

\subsection{A possible third order phase transition in 4D gluodynamics} 
We revisited the question of the convergence of lattice perturbation theory 
for a pure $SU(3)$ lattice gauge theory in 4 dimensions.
Using the most recent calculation of the 
weak coupling expansion of the plaquette average, we showed that the extrapolated ratio and the extrapolated slope 
suggest a nonanalytical power behavior at $\beta=6/g^2\simeq5.7$ with an exponent $\gamma \simeq -1.1$ 
in agreement with an existing analysis \cite{rakow2002}.
We found indications for a possible singularity in the third derivative of the free energy 
on $6^4$ and $8^4$ lattices.
As the lattice size increases, the statistical errors become large and a significantly larger 
number of independent configurations is needed in order to draw definite conclusions. This will be discussed 
in a forthcoming preprint \cite{lilipro}.

\subsection{A proposal for a ``perfect" field cut in Lattice gauge perturbation theory} 
We considered the effects of a field cutoff on the weak coupling series of a one plaquette $SU(2)$ lattice gauge theory. It possible to pick a the (perfect) field cutoff in such a way that the series {\it 
converges} toward the {\it correct} answer. We are considering the implementation of the method with 
a Langevin equation and its extension for four dimensional lattice gauge theory. This will be discussed 
in a forthcoming preprint \cite{lilipro}.

\acknowledgments
We thank D. K. Sinclair for making this workshop possible. We thank the participants of the workshop and 
A. Gonzalez-Arroyo, M. Creutz, F. Di Renzo and P. van Baal for valuable conversations.


\begin{thebibliography}{19}
\expandafter\ifx\csname natexlab\endcsname\relax\def\natexlab#1{#1}\fi
\expandafter\ifx\csname bibnamefont\endcsname\relax
  \def\bibnamefont#1{#1}\fi
\expandafter\ifx\csname bibfnamefont\endcsname\relax
  \def\bibfnamefont#1{#1}\fi
\expandafter\ifx\csname citenamefont\endcsname\relax
  \def\citenamefont#1{#1}\fi
\expandafter\ifx\csname url\endcsname\relax
  \def\url#1{\texttt{#1}}\fi
\expandafter\ifx\csname urlprefix\endcsname\relax\def\urlprefix{URL }\fi
\providecommand{\bibinfo}[2]{#2}
\providecommand{\eprint}[2][]{\url{#2}}

\bibitem{lat2004}
L.~Li and Y.~Meurice, Effects of large field cutoffs in scalar and gauge models, hep-lat/0409096 (to appear in the Lattice 2004.
Proceedings).
\bibitem[{\citenamefont{Meurice}(2002)}]{convpert}
\bibinfo{author}{\bibfnamefont{Y.}~\bibnamefont{Meurice}},
  \bibinfo{journal}{Phys. Rev. Lett.} \textbf{\bibinfo{volume}{88}},
  \bibinfo{pages}{141601} (\bibinfo{year}{2002}), \eprint{hep-th/0103134}.

\bibitem[{\citenamefont{Kessler et~al.}(2004)\citenamefont{Kessler, Li, and
  Meurice}}]{optim}
\bibinfo{author}{\bibfnamefont{B.}~\bibnamefont{Kessler}},
  \bibinfo{author}{\bibfnamefont{L.}~\bibnamefont{Li}}, \bibnamefont{and}
  \bibinfo{author}{\bibfnamefont{Y.}~\bibnamefont{Meurice}},
  \bibinfo{journal}{Phys. Rev.} \textbf{\bibinfo{volume}{D69}},
  \bibinfo{pages}{045014} (\bibinfo{year}{2004}), \eprint{hep-th/0309022}.

\bibitem{heller}
U. Heller and F. Karsch, Nucl. Phys. {\bf B 251} 254 (1986)

\bibitem[{\citenamefont{Alles et~al.}(1994)\citenamefont{Alles, Campostrini,
  Feo, and Panagopoulos}}]{alles93}
\bibinfo{author}{\bibfnamefont{B.}~\bibnamefont{Alles}},
  \bibinfo{author}{\bibfnamefont{M.}~\bibnamefont{Campostrini}},
  \bibinfo{author}{\bibfnamefont{A.}~\bibnamefont{Feo}}, \bibnamefont{and}
  \bibinfo{author}{\bibfnamefont{H.}~\bibnamefont{Panagopoulos}},
  \bibinfo{journal}{Phys. Lett.} \textbf{\bibinfo{volume}{B324}},
  \bibinfo{pages}{433} (\bibinfo{year}{1994}), \eprint{hep-lat/9306001}.
  
\bibitem[{\citenamefont{Di~Renzo et~al.}(1995)\citenamefont{Di~Renzo, Onofri,
  and Marchesini}}]{direnzo95}
\bibinfo{author}{\bibfnamefont{F.}~\bibnamefont{Di~Renzo}},
  \bibinfo{author}{\bibfnamefont{E.}~\bibnamefont{Onofri}}, \bibnamefont{and}
  \bibinfo{author}{\bibfnamefont{G.}~\bibnamefont{Marchesini}},
  \bibinfo{journal}{Nucl. Phys.} \textbf{\bibinfo{volume}{B457}},
  \bibinfo{pages}{202} (\bibinfo{year}{1995}), \eprint{hep-th/9502095}.

\bibitem[{\citenamefont{Di~Renzo and Scorzato}(2001)}]{direnzo2000}
\bibinfo{author}{\bibfnamefont{F.}~\bibnamefont{Di~Renzo}} \bibnamefont{and}
  \bibinfo{author}{\bibfnamefont{L.}~\bibnamefont{Scorzato}},
  \bibinfo{journal}{JHEP} \textbf{\bibinfo{volume}{10}}, \bibinfo{pages}{038}
  (\bibinfo{year}{2001}), \eprint{hep-lat/0011067}.
\bibitem{mo}
M. Ogilvie, this workshop; P. Meisinger and M. Ogilvie, hep-lat/0409136, (to appear in the Lattice 2004
Proceedings).
\bibitem[{\citenamefont{Balian et~al.}(1979)\citenamefont{Balian, Drouffe, and
  Itzykson}}]{balian74err}
\bibinfo{author}{\bibfnamefont{R.}~\bibnamefont{Balian}},
  \bibinfo{author}{\bibfnamefont{J.~M.} \bibnamefont{Drouffe}},
  \bibnamefont{and} \bibinfo{author}{\bibfnamefont{C.}~\bibnamefont{Itzykson}},
  \bibinfo{journal}{Phys. Rev.} \textbf{\bibinfo{volume}{D19}},
  \bibinfo{pages}{2514} (\bibinfo{year}{1979}).
  \bibitem[{\citenamefont{Li and Meurice}()}]{lilipro}
\bibinfo{author}{\bibfnamefont{L.}~\bibnamefont{Li}} \bibnamefont{and}
  \bibinfo{author}{\bibfnamefont{Y.}~\bibnamefont{Meurice}}, \bibinfo{note}{in
  preparation}.

  \bibitem[{\citenamefont{Kogut}(1980)}]{kogut80}
\bibinfo{author}{\bibfnamefont{J.~B.} \bibnamefont{Kogut}},
  \bibinfo{journal}{Phys. Rept.} \textbf{\bibinfo{volume}{67}},
  \bibinfo{pages}{67} (\bibinfo{year}{1980}).
\bibitem{negbet}
L.~Li and Y.~Meurice, Gluodynamics at negative $g^2$, hep-lat/0410029.

\bibitem[{\citenamefont{Bender and Boettcher}(1998)}]{bender}
\bibinfo{author}{\bibfnamefont{C.~M.} \bibnamefont{Bender}} \bibnamefont{and}
  \bibinfo{author}{\bibfnamefont{S.}~\bibnamefont{Boettcher}},
  \bibinfo{journal}{Phys. Rev. Lett.} \textbf{\bibinfo{volume}{80}}
  (\bibinfo{year}{1998}).
\bibitem[{\citenamefont{Horsley et~al.}(2002)\citenamefont{Horsley, Rakow, and
  Schierholz}}]{rakow2002}
\bibinfo{author}{\bibfnamefont{R.}~\bibnamefont{Horsley}},
  \bibinfo{author}{\bibfnamefont{P.~E.~L.} \bibnamefont{Rakow}},
  \bibnamefont{and}
  \bibinfo{author}{\bibfnamefont{G.}~\bibnamefont{Schierholz}},
  \bibinfo{journal}{Nucl. Phys. Proc. Suppl.} \textbf{\bibinfo{volume}{106}},
  \bibinfo{pages}{870} (\bibinfo{year}{2002}), \eprint{hep-lat/0110210}.
  


\end{thebibliography}
\end{document}